Paper presented at IATC (CATS) Institut Teknologi Maju UPM 2005,
6-8 December, IOI Marriott, Putrajaya, Malaysia.

# "Dynamic Phase Structures in the Evolution of Conventions" [a]

by

A.K.M. Azhar[b], WAT Wan Abdullah[c], and M. Ruzlan[d]

**Abstract**

This paper describes an agent-based model of a finite group of agents in a single population who each choose which convention to advocate, and which convention to practice. Influences or dependencies in agents choice exists in the form of "guru effects" and "what others practice". With payoffs being dependent on cumulative rewards or actual standings in society, we illustrate the evolutionary dynamics of the phase structure of each group in the population via simulations.

---

[a] Financial Support from Universiti Putra Malaysia Fundamental Research Grant #55171 is gratefully acknowledged.

[b] Graduate School of Management, Universiti Putra Malaysia 43400, Serdang, Selangor, Malaysia. Email: akmazhar@putra.upm.edu.my

[c] Department of Physics, Universiti Malaya 50603, Kuala Lumpur, Malaysia. Email: wat@um.edu.my

[d] Department of Industrial Technology, UDM, 21300 Terengganu, Malaysia. Email: mruzlan@udm.edu.my

## 1. Introduction

Conventions are usually the behavioural values or norms to which the majority of the population, society or community follows, and if a convention is part of a society, usually it is to the benefit of its members to conform [1], [2]. Overtime, conventions evolve; some goes into conflict, some became stronger, some decay, some may just disappear, some may get absorbed by a larger one, or some may even co-exist [2], [10].

Emergence, growth, decay, co-existence, or disappearance of conventions can be due to many causes or combinations of factors and influences. History or path dependence [8] and culture [5] could be a major factor. Technological development [9] may also be one. For example, at one time we hear affluence in a community as being attributed with the number of coconut trees and monkeys owned. However, technological breakthrough rendered this value obsolete. Sometimes we have the situation where a community's values and needs are modest, and joining a particular society will then be due to the need of being modest –possibly, we then have the case of the frog choosing to jump into the pond full of smaller ones [9].

Contrast the above scenario with an example of the subsequent emergence and decay of a convention. We have this community whose members are experiencing bigger desires, ambitions, or needs. In the positive stance yes, as members seek to improve their individual lot, fair competitive values emerge in their society generating positive welfare spillovers – however when these desires become *heinous* [6] - we then have the decay of the earlier convention.

Lack of knowledge, scared, *playing safe for fear of rocking the boat* [6] or whatsoever reasons may subsequently cause one's advocate of any particular value or stand to be heavily influenced by what others favour in society. Communal influence or *guru effects* is said to play a significant part in conventional formation. Furthermore we may also have occurrences whereby in the zeal of obtaining the end-rewards or payoffs in society, the particular values or stand that one adopts may not even be similar to that which is being advocated. This is the case of where profess and practice seems to belong to two entirely different spheres! and we observe the emergence of hypocrisy in the system.

Within the same society we may also have occurrences where the type of convention that is being adopted is being influenced not only by the type of convention which others practice but also by what others advocate. It is when a particular society has the above three dependencies evolving in their community that forms the framework of this paper. The objective of this paper is to model the consequences of these dependencies upon agents' choice, and to illustrate the evolutionary dynamics of the proportions of each group via simulation.

## 2. The Model:

We have a society or institution consisting of agents $i$ who choose which convention to favour and which to join.

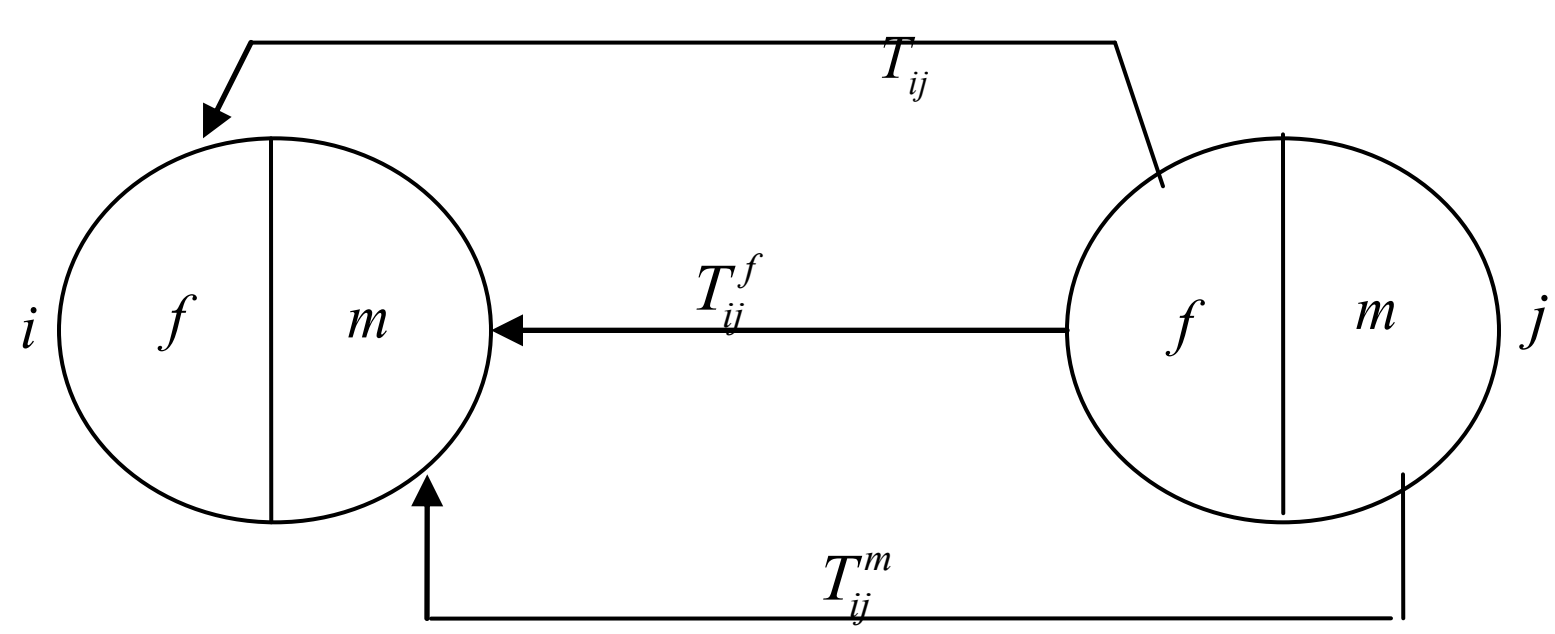

Figure 1

**A. Choice of Convention to Favour (** $f$ **)**: In this society, let's say each agent $i$ favours a convention: ( $f_i = \pm 1$ ). However, the type of convention which agent $i$ advocates is being influenced by the sum of what other agents $j$ advocate i.e. $f_i$, is parameterized by $T_{ij}$, the adaptive parameter represents the guru effects or influence of $j$ on $i$. We have $f_i \to +1$ if $\sum_j T_{ij} f_j \ > rand[-\sigma, \sigma]$, $\sigma$ equals noise level, and $rand$ = random number between given limits

$f_i \to -1$ →-1 otherwise

i.e. Agent $i$ choice is +1 if total influence of $j$ for +1 greater than for -1, and vice-versa.

The above set-up $f_i$ is similar to a Hopfield-like neural network model [7], and is stable if $T_{ij}$ is symmetric. There's also some possibility of it being chaotic if it is not stable [11]. However since $f_i$ is stable, we attempt to make the dynamics more interesting by differentiating between $f_i$ and $m_i$.

**B. Choice of Convention to Practice ($m$)** Each agent $i$ chooses to practice a convention: ($m_i = \pm 1$). The choice of convention which agent $i$ practice is being influenced by the sum of what other agents' $j$ advocate $(f_j)$ and practice $(m_j)$. $m_i$ is therefore parameterized by $T_{ij}^f$ and $T_{ij}^m$.

$m_i \to +1$ if $\sum_j (T_{ij}^f f_j + T_{ij}^m m_j) \ > rand[-\sigma, \sigma]$, $\sigma$ equals noise level, and $rand$ = random number between given limits

$m_i \to -1$ otherwise

Agent $i$ choice is +1 if total influence of $j$ for $c(+1)$ is greater than for -1, and vice-versa.

**C. Payoffs**

Each agent earns payoff $p_i = P(c_i) = P(m_i)$ for this choice of $m_i$ and the cumulative payoff of agent $i$ is given by $g_i \to g_i + p_i g$ which is a measure of 'standing' in society. Furthermore we have payoffs depending also on the amount of agents of good standing:

$P(c) = \sum_{i:f=c} g_i$ (sum of standings of those who advocate convention $c$). This is normalised and scaled so that $P(+1) + P(-1) = 1$

Learning – influence matrices are adjusted according to payoffs and gains in payoffs use linear Hebb-like [4) learning [3].]

$\Delta T_{ij} = \eta f_i f_j P(f_i)$ (depends on payoff in advocation)

$\Delta T_{ij}^{f} = \eta m_i f_i (P(m_i) - P(-m_i))$ (depends on gain in payoff in practice)

$\Delta T_{ij}^{m} = \eta m_i m_j (P(m_i) - P(-m_i))$

($\eta$ is the learning rate) (all are normalized)

We start the simulation with random initial values

## 3. Simulations:

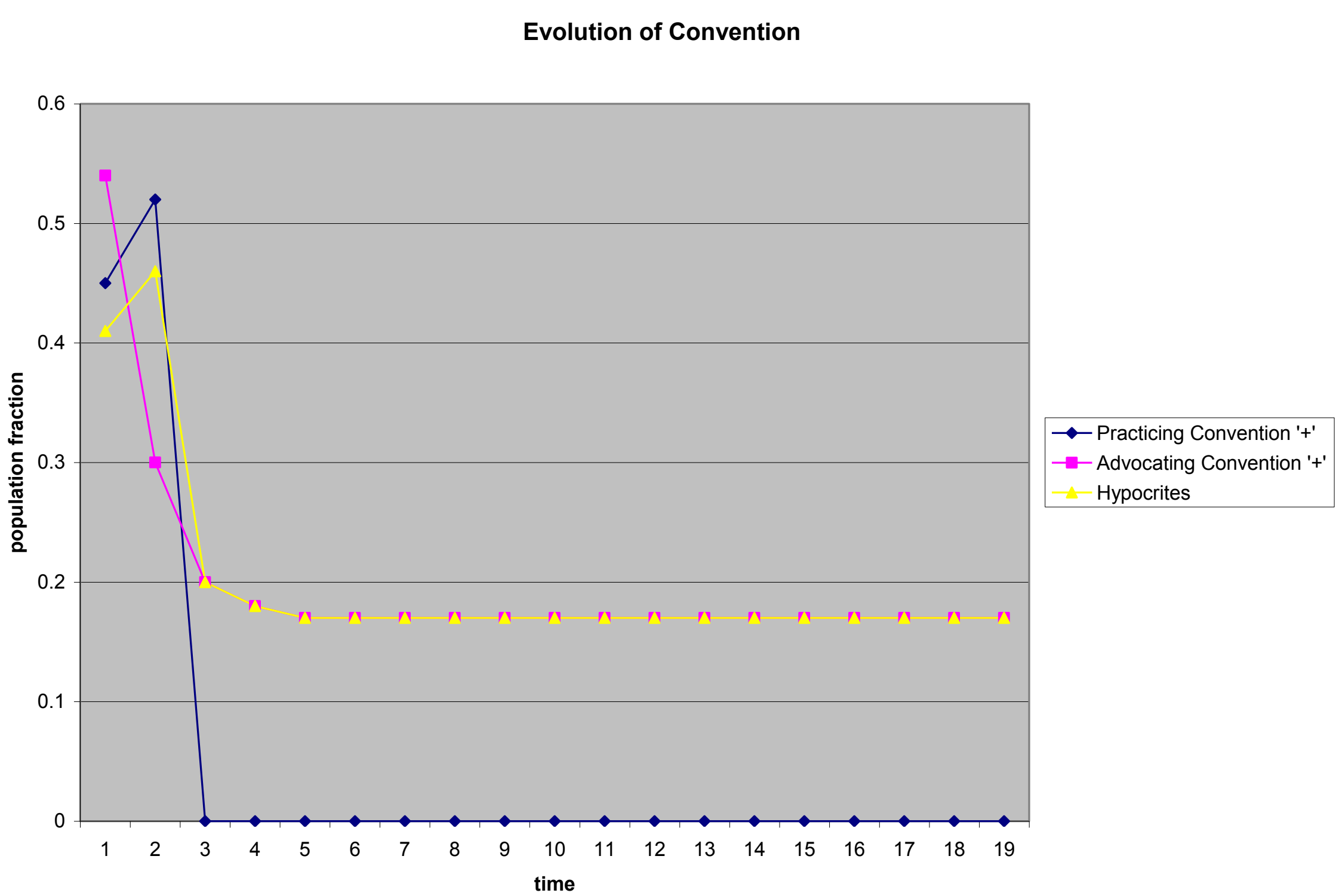

Observe that the dynamics is stable for some choice of parameters.

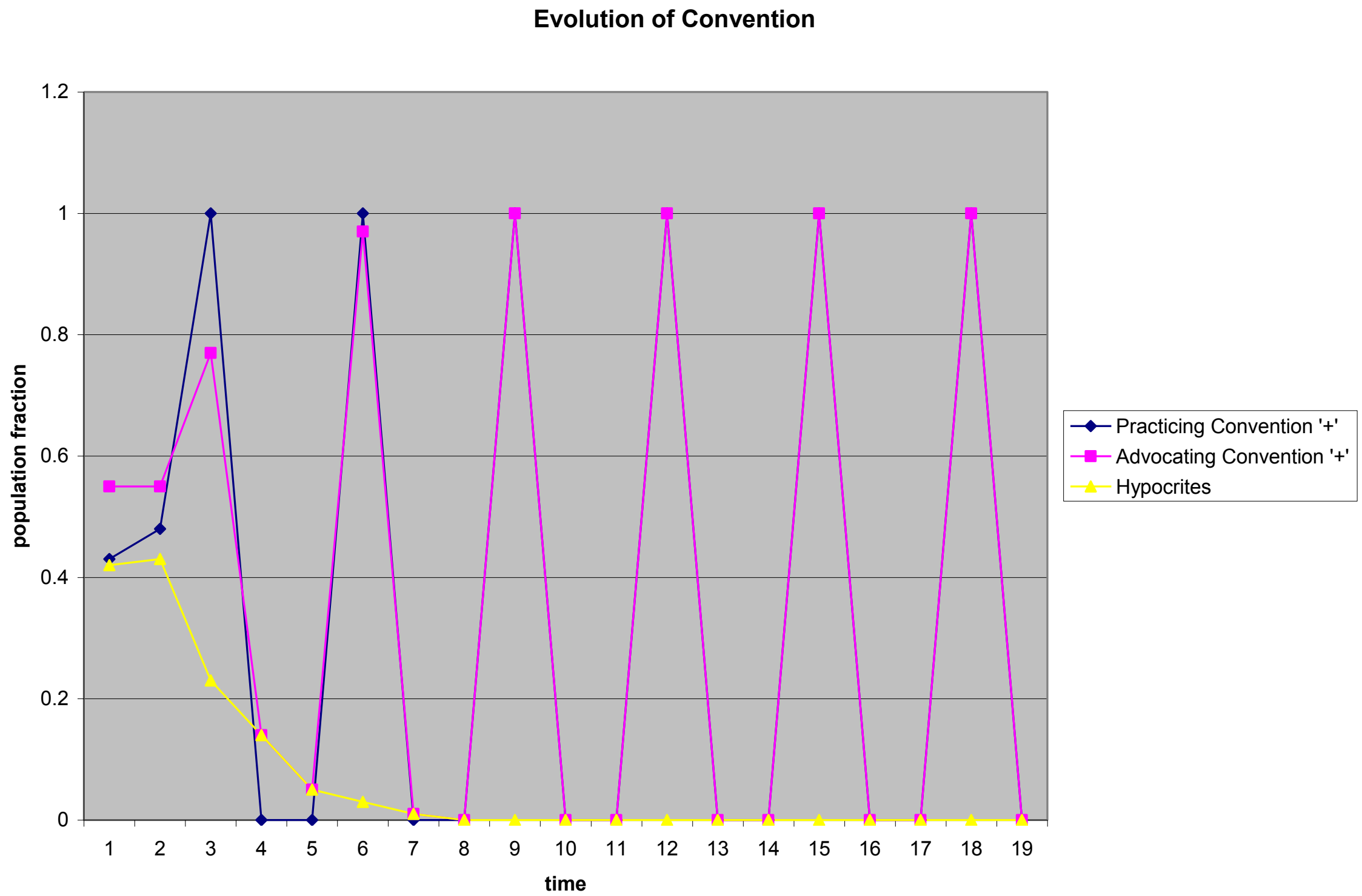

Observe that the dynamics is oscillatory for some choice of parameters.

The system a lot of times freezes into a single convention (despite the existence of the opposing view, arising in hypocrisy), but it may also exhibit cyclic behaviour as the figures above show.